\documentclass[
preprint,
 amsmath,amssymb,
 aps,
pra,
]{revtex4-2}
\usepackage{graphicx}
\usepackage{dcolumn}
\usepackage{bm}

\begin{document}

\title{Quantum learning machines. }

\author{Gerard  Milburn}
\email{milburn@physics.uq.edu.au}
\affiliation{ARC Centre of Excellence for Engineered Quantum Systems, School of Mathematics and Physics\\
 The University of Queensland, Brisbane, Australia 4072}

\date{\today}

\begin{abstract}
 Physical learning machines, be they classical or quantum, are necessarily dissipative systems. The rate of  energy dissipation decreases as the learning error rate decreases linking thermodynamic efficiency and learning efficiency.   In the classical case the energy is dissipated as heat.  We give an example based on a quantum optical perceptron where the energy is dissipated as spontaneous emission.  At optical frequencies the temperature is effectively zero so this perceptron is as efficient as it is possible to get. The example illustrates a general point: In a classical learning machine, measurement is taken to reveal objective facts about the world. In quantum learning machines what is learned is defined by the nature of the measurement itself.    
\end{abstract}

\maketitle



\section{\label{intro}Introduction}
Biology offers abundant evidence that physical systems can learn, that is to say, physical systems can exhibit stable behaviour, conditioned on prior interactions with an external environment, in order to achieve a goal. The existence of a goal does not imply teleology. A goal can arise spontaneously: In biological systems the goal is to survive long enough to reproduce. We are entering an era in which learning machines can be engineered. In which case, what are the physical principles in play? 

I want to make a distinction between learning {\em algorithms} and learning {\em machines}. Modern machine learning refers to a class of algorithms running on digital CMOS chips. On the other hand, a learning machine could be instantiated in any physical system and not necessarily digital. Biological learning is not based on algorithms running on digital computers\cite{Purves}, even if it can be simulated that way.  In this this paper I discuss the physical principles required for machines to learn.

A learning machine is an open, dissipative physical system driven far from thermal equilibrium by access to a low entropy source of energy, for example, a battery. I will focus on simple classification in supervised learning. Here the objective is to learn a binary valued function, $f(x)$ of the input data $x$ by giving the machine a list of examples $(x, f(x)$ and adjusting the parameters of the machine through feedback so that the actual output $\tilde{f}(x)$ has a low probability of error: $\tilde{f}(x)\neq f(x)$. The goal is do this while making efficient use of the available thermodynamic resources. I am interested in quantum machines operating at very low temperature, in which case the goal is to learn in the presence of unavoidable quantum noise. As will become clear, noise is essential for learning, so we need to ask how quantum noise might be harnessed for efficient learning. 

 \section{Physical perceptrons.}
Neural network machine learning {\em algorithms}  are based on the concatenation of elementary non linear functions called the activation functions. The preceptron function is an example\cite{Russell-Norvig}.  In the simplest examples, the input to the function is a linear combination of data elements and the coefficients are called weights. The output is a single real number, $\hat{y}$. The learning algorithm consists in evaluating the function over many examples of input data, labelled with a true output value $y$. The actual output $\hat{y}$ is compared to the true value and the weights adjusted for the next iteration in such a way (back propagation) as to cause the actual value to equal the true value most of the time. The key point is that as the network learns, the distribution over  weights change such that it becomes concentrated on particular points in weight-space with small fluctuations over different training episodes. The algorithm is typically implemented on a standard CMOS processor. 

In a physical perceptron, the activation function is replaced by a physical device that implements a noisy switch and the concatenation of functions is replaced by a physical network of activation switches using  feed-forward signals from the output of one layer of switches to the next. Learning consists in an actual feedback of signals structured to emulate back propagation. In this picture learning is best regarded as cooling in weight space\cite{CP-LM-review}.   The physical perceptron is a dissipative non linear dynamical system driven far from thermal equilibrium and subject to noise.

It is widely acknowledged that neural network algorithms running on standard CMOS devices consume large amounts of power both for learning and for inference\cite{Chojnnacka}. It is hoped that physical neural networks, based on physical perceptrons, could be more thermodynamically efficient. After all, the human brain is some kind of physical learning machine and consumes a tiny amount of power by comparison with deep learning technology. 

The noise in an activation switch is due to thermal fluctuations that necessarily accompany the dissipation required for the switch to function (the fluctuation-dissipation theorem\cite{Gardiner}).  In an elementary switching event the energy dissipated must be proportional to $k_B T$  where $k_B$ is Boltzmann's constant and $T$, the ambient temperature. The power dissipated will depend on the product of the switching rate and the local temperature.  

\begin{figure}
    \centering
    \includegraphics[scale=0.5]{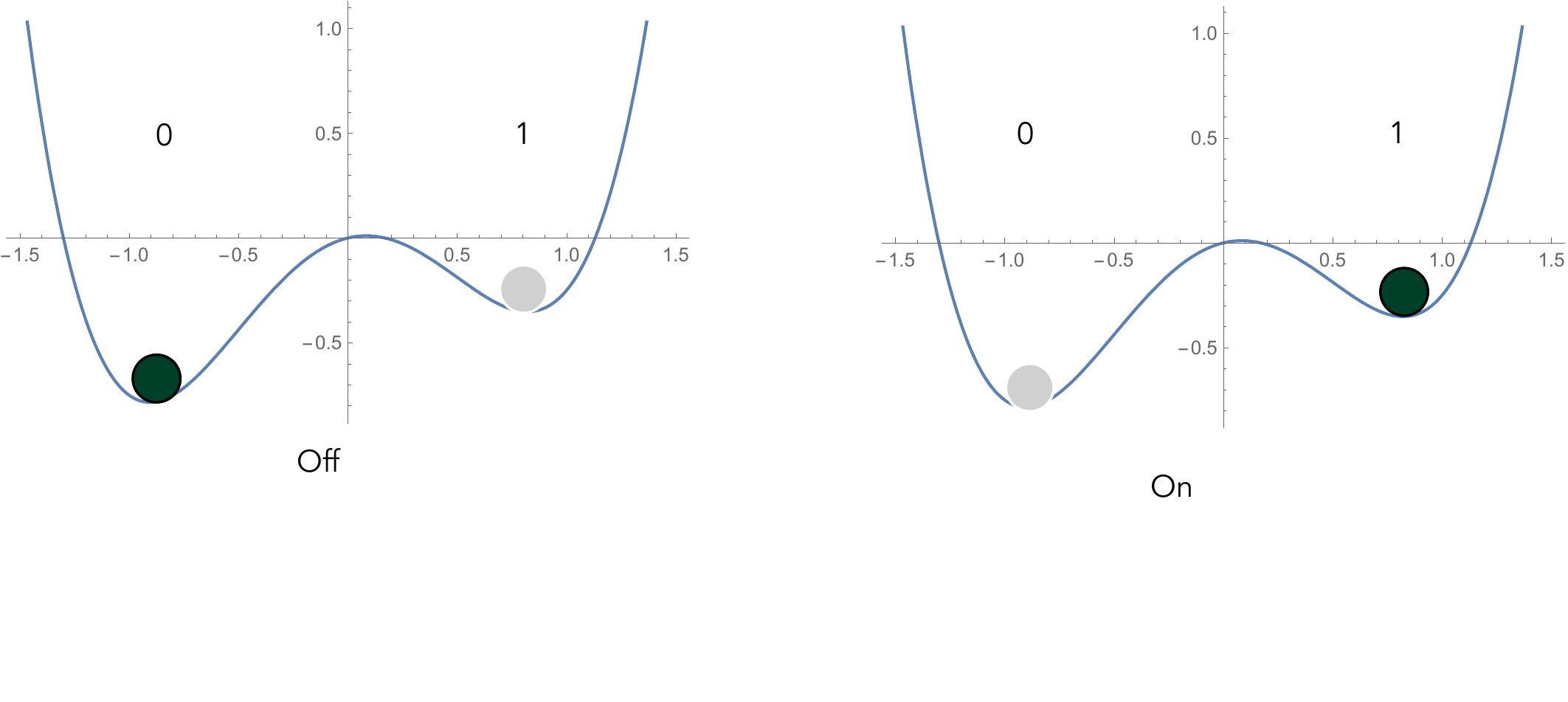}
    \caption{A physical activation switch modelled as a particle moving in a double well potential with very high friction.  On the left the switch is in the off configuration. As a linear potential is applied, a bias force does work on $W$ the particle until, at some random value for $W$, it undergoes a thermally activated transition to the other well and moves under friction to the new local equilibrium. It is now in the on configuration. }
    \label{double-well}
\end{figure}

To see this consider a simple a particle moving in a symmetric double well potential with high friction, see Fig. (\ref{double-well}).
Applying a time varying  bias force will make the double well asymmetric such that one well is deeper than the other. As the bias is slowly increased from zero, work is done on the system. At some value of the bias, and the work done, the thermal noise will cause the particle to overcome the potential barrier between the wells and jump to the lower well. As this is a stochastic process, the work done in that trail, $W$, is also a random variable. Using the Jarzynski equality\cite{Jarzynski} we can relate the average of $W$ over many trials to the change in free energy of the equilibrium steady state distributions between the starting  and stopping times of the bias variation,
\begin{equation}
    \langle e^{-W/k_BT}\rangle = e^{-\Delta F/k_BT}
\end{equation}
where $\Delta F $ is the change in the free energy between the initial and final equilibrium states. 

The work done is dissipated as heat after the transition as the particle moves to the new local equilibrium state. Heat $\delta Q$ flows into  the environment due to friction as the particle moves to bottom of the deeper well. The energy lost due to heat is proportional to $k_BT$ and one can reduce this cost by reducing the temperature. The problem  is that, at low temperatures, the switching rate also goes to zero. This is a consequence of Kramer's theory of first passage times\cite{Gardiner}. This would result in the learning rate of a perceptron network going to zero.

If we coarse-grain the output observable to be simply the sign of the output, the stochastic dynamics can be approximated by a two-state markov process\cite{CP-LM-review}. Let the state be specified by $n=+1$ if the particle is localised on the right and $n=-1$ if the particle is localised on the left.  A simple learning model based on a single double well perceptron can now be specified.

Suppose we want to train a perceptron to implement a one bit Boolean function. There are four such functions: two produce a constant output regardless of the input, one simply copies the input to output and the other inverts the input to produce the output. This last one is the NOT gate: $y_{out}=y_{in}\oplus 1$ where the binary operation is addition mod-2.  

The input $y\in \ { 0,1\ }$ is encoded in a linear bias potential with slope $x=2y-1$. The weights correspond to another a linear bias potential of arbitrary slope $w$ a real number. The total bias acting on the particle in the double well is then given by $A(w,x)= x.w$.  We set up the double well such that if $A>0$ the particle has a higher probability to transition to the right well, while if $A<0$ the particle has a higher probability to transition to the left well. Work is done and heat dissipated each time the bias potential changes, but each of these is a random variable as sometimes the transition does not take place. Typically the conditional probability to make the transition is given by thermal activation so that 
\begin{equation}
    p(w) = (1+e^{-\beta A(w)})^{-1}
\end{equation}
where $\beta =(k_BT)^{-1}$ with $k_B$ Boltzmann's constant and $T$ the temperature of the environment. This is a sigmoidal function of the weight.

The training set is a large number of ordered pairs of the form $x,n_T$  drawn at random from $\{ (-1,1), (1,-1)\}$. Initially the weights $w$ are drawn from a Gaussian distribution with mean and variance $\sigma$. In each single trial we compare the actual output of the switch $n$ to the label $n_T$ for that data point and compute the error $\epsilon= (n-n_T)^2/4$. We now define an {\em epoch} as a large number of runs using the same training data point and compute the average error $\bar{\epsilon}$ and average output $\bar{n}$ over the epoch. This is equivalent to sampling the switching probability. It is easy to see that
\begin{eqnarray}
   \bar{\epsilon}(w) & = & \frac{1}{2}(1-n_T \bar{n}(w))\\ 
   \bar{n}(w)& = & 2p(w)-1.
\end{eqnarray}

We now apply a feedback to change the weights for the next epoch. This is done in such a way as to decrease the average error and is known as gradient descent. The feedback rule is\cite{CP-LM-review} 
\begin{equation}
    \Delta w=\beta n_T x(1-\bar{n}^2(w))/4.
\end{equation}
for input $x,n_T$ at that epoch. The corresponding change in the average error is 
\begin{equation}
\Delta \bar{\epsilon}(w)=-\beta^2(1-\bar{n}^2(w))^2/16.
\end{equation} 
In the case of the NOT gate $n_Tx < 0$ for every training datum so that the weights always decrease. In this simple example this means if the weight $w$ is initially chosen as negative, the learning is very fast. 

If we focus on the stochastic dynamics of the weight $w$ we see that it is highly nonlinear process in which the initial broad Gaussian distribution, centered on $w=0$, converges to a much more narrow distribution centered on $w<0$. In physical terms this is cooling via feedback and is a general feature of all physical learning machines. The process of learning decreases the entropy of the machine while increasing the entropy of the environment by heat dissipation.  It can only occur if the learning machine is driven to a non-thermal equilibrium steady state through access to a low entropy source of energy.  This has very important consequences for our understanding of how learning machines arise in biological systems through evolution\cite{CP-LM-review}. The evolution of biological systems that learn is favoured by the laws of physics when access to thermodynamic resources determines survival.

At very low temperatures the Kramer's rate formula fails, but dissipative quantum tunnelling, also called quantum activation\cite{Dykman}, can drive the transitions in an activation switch in a bistable system. Learning machines will need to use quantum switching if they are to make the most of thermodynamic resources. We will discuss a different example drawn from quantum optics in which quantum noise in the form of spontaneous emission drives the transition.

\section{A quantum optical  perceptron at zero temperature.}
As an example of a quantum activation switch, and a quantum perceptron, we will consider Raman single-photon sources (RSPS) and Raman single-photon detectors (RSPD)\cite{KMS}. The ideal RSPS produces a single photon pulse with a controllable temporal mode shape. The ideal RSPD is a single-photon detector which projects a single-photon pulse onto a re-configurable temporal mode. A classical write pulse (RSPS) and a classical read pulse (RSPD) are used to control temporal mode shapes.   This level of control can be used to encode information in a single photon  and change the weights for a single photon detection. 

In normal operation, the output of the RSPD is a single binary number corresponding to the absorption of the single photon or reflection of the single photon from the detector. The latter event signals an error and can be detected using a standard single photon counter and the record is a classical random variable, as is the case of a classical activation switch. However it would be better if the output of a quantum perception was the same kind of quantum object as the input, i.e. a temporally encoded single photon. In that way an all-optical feed forward neural network could be constructed. 

\begin{figure}
    \centering
    \includegraphics[scale=0.5]{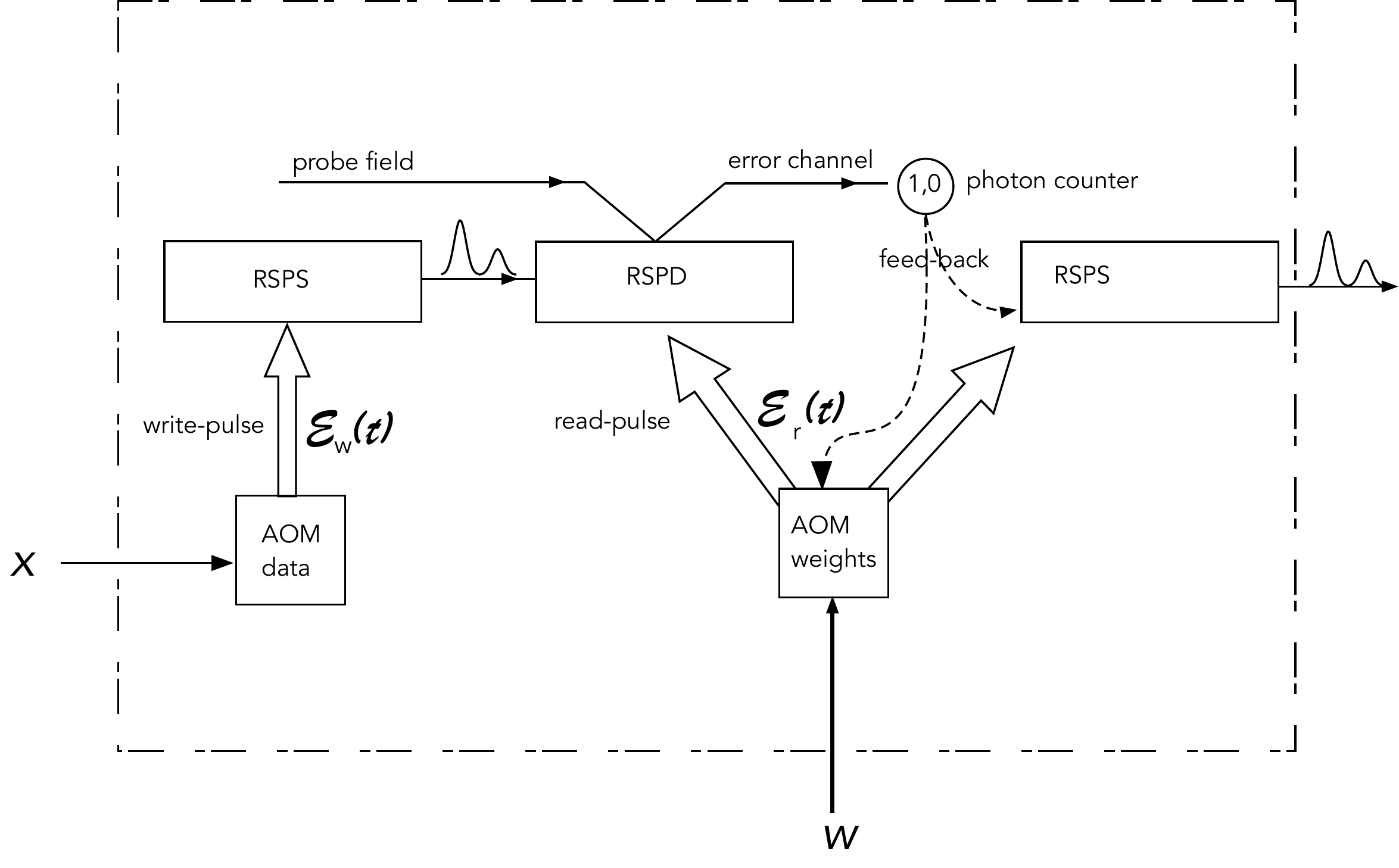}
    \caption{Data is encoded into the temporal shape of a Raman single-photon source (RSPS) using data to amplitude modulate the write-field.  The photon is injected into a Raman single-photon detector (RSPD) using a read-field that is controlled by feedback amplitude modulation. This determines the weights used by the preceptron. When the read-field matches the write-field the photon is absorbed with high probability in the RSPD, otherwise it is reflected and lost. Absorption is by photo counting. This resulting classical signal is used to control the weights in the write field until the error is a minimum. Then it is used to switch the RSPD to a RSPS where upon the write field generates a photon closely matching the data encoded photon.}
    \label{Raman-perceptron}
\end{figure}

This is easily achieved. Once the write field has effectively stored a photon of an unknown temporal shape, the electronic state of the RSPD can be swapped turning it into a RSPS. Now the control field will produce a single-photon pulse with the same shape as the learned shape, with high probability. The scheme is indicated in Fig. (\ref{Raman-perceptron}).   This is fine for a single perceptron, but in a neural network we need to consider the case of multi-photon inputs to the Raman memory. That is beyond the scope of this paper.  

\begin{figure}
    \centering
    \includegraphics[scale=0.5]{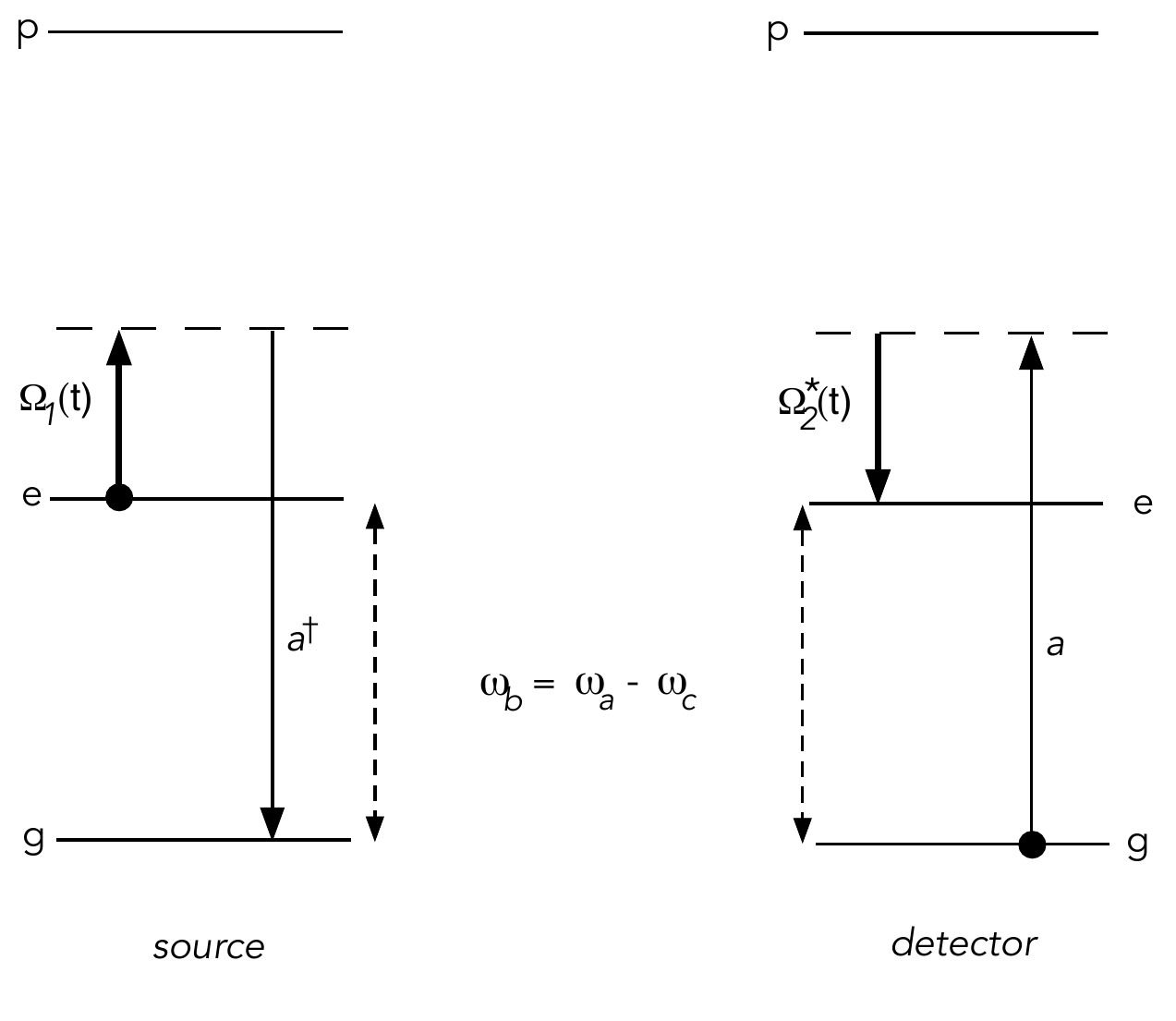}
    \caption{A representation of the Raman transition for a single-photon source (left) and single-photon detector (right). An atomic transition is coupled to an optical cavity field (frequency $\omega_a$) by a Raman process in which a classical control field is tuned to the frequency difference of the atomic system and the cavity, $\omega_c=\omega_a-\omega_b$.  The classical control field amplitude is  ${\cal E}_j(t)$ where $j=1$ corresponds to the write-field, and $j=2$ corresponds to the read-field. In the case of the source (left), the atom is prepared in the excited state. In the case of a detector, (right) the atom is prepared in the ground state. The atomic state labelled $p$ is a probe state to enable a dispersive read-out of the occupation of states $g,e$.  }
    \label{raman-scheme}
\end{figure}

A single atom in an optical cavity,  with a time dependent pump tuned to a Raman resonance, can be configured to realise a single-photon source or  detector depending on the initial state of the atom\cite{james_atomic-vapor-based_2002}.  Let the frequency of the optical cavity be $\omega_a$ and the frequency of the lower transition be $\omega_b$. We then take the carrier frequency of the classical control field as $\omega_c=\omega_a-\omega_b$ as the Raman resonance condition.  This enables a single photon source or detector depending on how the Raman system is prepared, see Fig. (\ref{raman-scheme}).  We will assume that the cavity mode is heavily damped so that the single-photon emission/absorption is fast. This means we can adiabatically eliminate the cavity field dynamics.  

In the case of a RSPS we prepare the atom in the ground state $g$.  The output mean photon emission rate from the cavity in the adiabatic limit is\cite{Ryan}
\begin{equation}
\langle a^\dagger_{out}a_{out}\rangle = |{\cal E}_w(t)|^2
\end{equation}
$|{\cal E}_w(t)|^2$ has units of  s$^{-1}$, so that this is the intensity of the external control field. 

The cavity emits a single photon excitation of the vacuum of the from
\begin{equation}
    |\nu(t)\rangle = \int_{-\infty}^\infty d\omega\ \tilde{\nu}(\omega) a^\dagger(\omega) |0\rangle 
\end{equation} where $a(\omega),a^\dagger(\omega)$ are the usual bosonic annihilation and creation operators. This is a coherent superposition of a single excitation over many frequency modes weighted with the complex amplitude function $\tilde{\nu}(\omega)$. In an ideal RSPS,  $\nu(t)={\cal E}_w(t)$ is the Fourier transform of $\tilde{\nu}(\omega)$. Thus we can control the temporal state of the emitted photon by an appropriation modulation of the control field.

In the case of a RSPD, we prepare the atom in the ground state $g$ and an itinerant single-photon pulse is absorbed by the cavity if the control field, the read-field, has the right temporal shape. If the photon is not absorbed, it is reflected from the cavity. We design the read-field to optimally absorb a single photon pulse into the atom-cavity system from the external field.

 If the input field of the cavity is prepared in a single-photon pure state with temporal amplitude $\nu(t)$, and the carrier frequency equal to the cavity frequency, the probability for absorption is simply the probability for the atom to make a transition $g\rightarrow e$. This is\cite{Ryan} 
\begin{equation}
\label{detect-prob}
p_A(t)= \left | \int_{-\infty}^t \  dt' {\cal E}_r^*(t)\nu(t')  e^{-(\tau(t)-\tau(t')/2}\right |^2
\end{equation}
where we define a dimensionless time variable, $\tau$ through the change of variable 
\begin{equation}
\tau(t) = \int_{-\infty}^t\ dt' |{\cal E}_r(t')|^2 
\end{equation}
and ${\cal E}_r(t)$ is the read-field amplitude. 
This clearly depends on the  time over which the read-field ${\cal E}_r(t)$ is  non zero. This is the detection time. In the long-time limit $p_1(t)$ tends to a constant less than unity.

\begin{figure}
    \centering
    \includegraphics[scale=0.5]{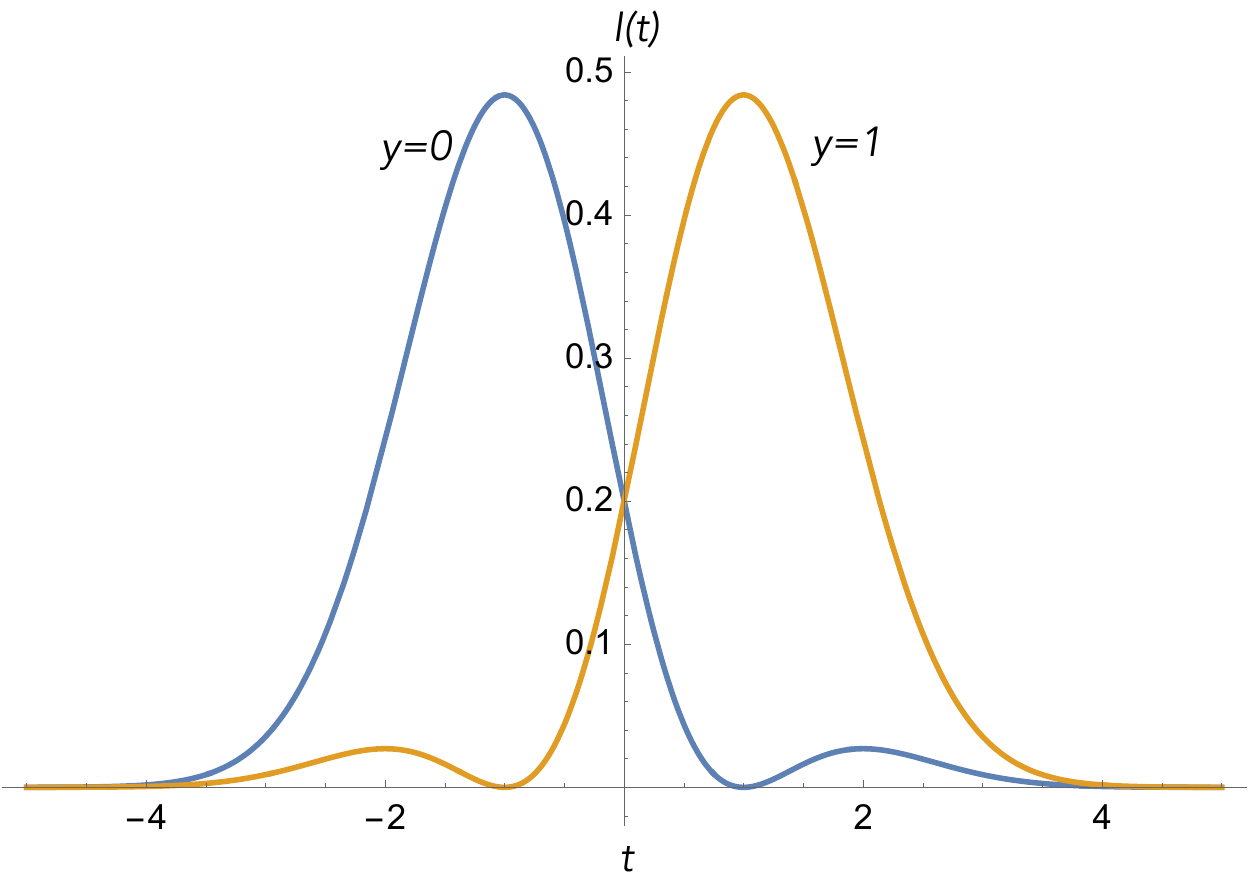}
    \caption{Pulse-code modulation encoding of a single binary variable in a single-photon state.  The perceptron learning a NOT gate is trained to reverse the  temporal order of the pulses.  }
    \label{time-bin}
\end{figure}

As an example we will encode the input $y$ in a pulse-code single-photon state. With respect to a reference time $t=0$, if $y=0$, the pulse intensity is a maximum for  $t<0$ while for $y=1$ the pulse intensity is a maximum for $t>0$ see Fig.(\ref{time-bin}). The NOT gate corresponds to reversing the times of the peak intensity. 
'

The modulation required in the single-photon source  is defined in terms of a superposition of Hermite Gaussian modes\cite{RayWal}, 
\begin{equation} 
\nu_y(t) =\frac{1}{\sqrt{2}}(u_0(t)+xu_1(t))
\end{equation}
where we define $x=2y-1$. The probability for the perceptron to switch is given by Eq. (\ref{detect-prob}) with the read-field  at the detector defined by
\begin{equation} 
{\cal E}_r(w,t) =\frac{1}{\sqrt{2}}(u_0(t)+e^{i\pi A(w)/2}u_1(t))
\end{equation}
where $w\in{\mathbb R}$ and the activation is $A(w)=xw-1$. 

As we approach $w=-1$ the device will approximate the NOT function. In that case,  if $x =-1 $ then $A(w)=0$ and the output photon has mode function $\nu_1(t)$. If  $x=+1$ then  $A(w)=1$ the output photon state is $\nu_0(t)$.

\section{Conclusion}
Both classical and quantum learning machines are irreversible dissipative devices. In the classical case the energy dissipated is called heat. In the quantum case it is simply spontaneous emission from a bosonic oscillator. In both cases, as the machine learns, the energy dissipated per trial decreases as the error decreases. 

In the photonic example, the energy is dissipated by photons lost from the system.  The loss of energy is one quanta of photonic energy and does not depend on temperature.  As the system learns, fewer and fewer trials lead to the relfection of the photon from the detector and the average energy lost per trial falls to a minimum. This is probably as energy efficient as it is possible to get at room temperature. 

In a full quantum photonic learning machine, there are as many photons as their are input channels however the photons become redistributed across the network in a coherent way. This is reminiscent of boson sampling where the Raman sources and detectors are simply replaced by beam-splitters. The optical scheme we have described is a generalisation of boson sampling to a  photonic quantum neural network. More work is needed to determine if this will provide a quantum advantage beyond the improvements in thermodynamic efficiency. 

The nature of the Raman detection scheme determines what is measured and  plays an essential role in what is learned.  Suppose we had used a conventional photon counter\cite{WallsMilburn}. In any trial, one and only one photon is counted and the detection times are distributed according to  $|\nu(t)|^2$. What we really want to measure is the temporal mode shape of the photon $\nu(t)$, in general this is a complex valued function of time. This is precisely what Eq. (\ref{detect-prob}) depends on. The Raman scheme enables one to sample the probability distribution that a single photon is in a chosen temporal mode function. In effect, what we require is a projection valued measure onto a class of single-photon states $P_{\mu}(t) dt=|\mu(t)\rangle\langle \mu(t) | dt $ such that the probability that a single photon in state $|\nu(t)\rangle$ will be found in state $|\mu(t)\rangle$ is 
\begin{equation}
P(\mu|\nu)= \left |\int_{-\infty}^\infty dt\ \langle \mu(t)|\nu(t)\rangle \right |^2 
\end{equation}
This is what the Raman detection scheme approximates. By variation over the class $|\mu(t)\rangle$ we can learn the temporal mode shape of a photon. In a classical learning machine measurement simply reveals objective facts about the world. In quantum learning machines what is learned is defined by the nature of the measurement itself.  This is a restatement of Bohr's complementarity for quantum learning machines.

 \section*{Acknowledgements}
 I wish to thank Sally Shrapnel, Sahar Basiri-Esfahani, Andrew Briggs, Michael Kewming and Ian Walmsley for fruitful discussions.  This research was supported by the Australian Research Council Centre of Excellence for Engineered Quantum Systems (EQUS, CE170100009). 

\bibliography{learning}

\end{document}